\documentclass[twocolumn,showpacs,aps,pra]{revtex4}

\usepackage{epsfig}
\usepackage{eufrak}

\begin{document}

\def\4he{$^4$He}
\def\pc{\protect\cite}
\def\hpsi{\hat\psi}
\def\tpsi{\tilde\psi}
\def\br{{\bf r}}
\def\bk{{\bf k}}
\def\bu{{\bf u}}
\def\bw{{\bf w}}
\def\brt{\br,t}
\def\bbrt{(\brt)}
\def\cphio{\Phi_0}
\def\beq{\begin{equation}}
\def\eeq{\end{equation}}
\def\bea{\begin{eqnarray}}
\def\eea{\end{eqnarray}}
\def\bna{\bbox{\nabla}}
\def\bp{{\bf p}}
\def\bv{{\bf v}}
\def\tn{\tilde n}
\def\tp{\tilde p}
\def\be{\bbox{\eta}}
\def\imag{{\rm{Im}}}
\newcommand{\eq}[1]{{(\ref{eq#1})}}
\newcommand{\tb}[1]{\textcolor{blue}{#1}}
\newcommand{\tr}[1]{\textcolor{red}{#1}}
\newcommand{\tg}[1]{\textcolor{green}{#1}}

\title{Damped Bogoliubov excitations of a condensate interacting with a
static thermal cloud} 

\author{J. E. Williams and A. Griffin} 

\affiliation{Department of Physics, University of Toronto, Toronto,
Ontario M5S 1A7, Canada} 

\date{\today} 

\begin{abstract}
We calculate the damping of condensate collective excitations at
finite temperatures arising from the lack of equilibrium between the
condensate and thermal atoms. Since a static thermal cloud already
produces damping in our model, we ignore the non-condensate dynamics.
We derive a set of generalized Bogoliubov equations for finite
temperatures that contain an explicit damping term due to collisional
exchange of atoms between the two components. We have numerically
solved these Bogoliubov equations to obtain the temperature dependence
of the damping of the condensate modes in a harmonic trap.  We compare
these results with our recent work based on the Thomas-Fermi
approximation.

\end{abstract}


\maketitle

\section{Introduction}

In this paper, we calculate the damping at finite temperatures of
collisionless condensate collective modes due to interactions with a
static thermal cloud.  Our treatment starts with the general theory
developed in Ref.~\cite{Zaremba99} and is an extension of the static
Hartree-Fock-Bogoliubov Popov (HFB-Popov)
approximation~\cite{Hutchinson97,Dodd98,Dodd98b} to include the effect
of the lack of diffusive equilibrium between the condensate and
non-condensate. The static thermal cloud acts as a reservoir that can
exchange particles with the condensate so as to bring the condensate
back into equilibrium, resulting in a damping of the collective
oscillations of the condensate. We do not expect, in general, that the
collisional damping we find from a static thermal cloud will be much
modified when we treat the non-condensate dynamically. Because we are
treating the non-condensate statically, Landau and Beliaev damping are
not contained within in our description.

Our present work is the natural extension of a recent
study~\cite{Williams2000} of the damping of collisionless condensate
collective modes at finite temperatures. This was based on a
simplified model treating the time-dependent Gross-Pitaevskii equation
for the condensate within the Thomas-Fermi approximation (TFA).  We
described the damping in terms of the density $\delta n_c(\br,t)$ and
velocity $\delta \bv_c(\br,t)$ fluctuations of the condensate
oscillations. At $T=0$, where the non-condensate component is absent,
Stringari~\cite{Stringari96} showed that a simple analytic solution of
the condensate collective modes could be obtained within the
Thomas-Fermi approximation. In Ref.~\cite{Williams2000}, we formally
generalized these ``Stringari'' solutions to finite temperatures by
assuming that the mean field of the static thermal cloud has a
negligible effect on the condensate oscillations. We then calculated
the damping due to the lack of equilibrium between the condensate and
non-condensate to first order in perturbation theory, using the
Stringari modes as a zeroth-order solution. In
Ref.~\cite{Williams2000} we compared this damping directly to Landau
damping and found it gives a $30\%$ to $50\%$ correction to the total
damping for typical experimental parameters.

The main focus of the present paper is to generalize our earlier
work~\cite{Williams2000} to include the same inter-component
collisional damping mechanism directly in the Bogoliubov equations,
still treating the non-condensate statically. This allows us to check
the range of validity of the TFA made in Ref.~\cite{Williams2000}. The
TFA is valid when the number of condensate atoms $N_c$ is large, that
is, when the condensate radius $R_{\rm{TF}}$ is much larger than the
harmonic oscillator length $r_{\rm{HO}}$~\cite{Dalfovo99} \beq
r_{\rm{HO}}/R_{\rm{TF}} = (15 N_c a/r_{\rm{HO}})^{-1/5} \ll 1, 
\label{eqTFA} \eeq where $a$ is the $s$-wave scattering length. 
The TFA will break down as one approaches the BEC transition
temperature $T_{\rm{BEC}}$ where $N_c$ becomes small. It is in this
temperature region where we must solve the coupled Bogoliubov
equations for our model, as we do in this paper. From the
time-dependent GP equation describing the fluctuations of the
condensate order parameter, $\delta \Phi(\br,t)$, we derive
generalized Bogoliubov equations for the collective mode amplitudes
$u_j(\br)$ and $v_j(\br)$ and complex energies $\varepsilon_j$, which
we solve numerically and compare to the predictions of our earlier TFA
calculation~\cite{Williams2000}. We also give a careful discussion of
the formal properties of coupled Bogoliubov equations which include
damping, since this may be of more general interest than our specific
model calculation.

\section{Static Popov approximation}

Our starting point is the finite $T$ generalized GP equation 
derived in Ref.~\cite{Zaremba99} (see also Refs.~\cite{Stoof99,Walser99}.)
\bea 
i\hbar {\partial \Phi(\br,t) \over \partial t}
&=& \Big [ - {\hbar^2 \over 2 m} \nabla^2 + U_{\rm{ext}}(\br) + g
n_c(\br,t) + 2g \tilde{n}(\br,t) \nonumber \\&-& i R(\br,t) \Big
] \Phi(\br,t) , 
\label{eq1} 
\eea 
where the interaction parameter $g = 4 \pi \hbar^2 a/m$, $a$ is the
$s$-wave scattering length, $n_c(\br,t) = |\Phi(\br,t)|^2$, and
$\tilde{n}(\br,t)=\int d\bp f(\bp,\br,t)/(2 \pi \hbar)^3$ is the 
non-condensate local density. The term $R(\br,t)$ in \eq{1} describes 
the exchange of atoms between the condensate and normal gas and is 
given by
\beq 
R(\br,t) = {\hbar \over {2 n_c(\br,t)}}\int {d\bp \over (2 \pi \hbar)^3}
C_{12}[f(\bp,\br,t),\Phi(\br,t)].
\label{eq2}
\eeq 
This involves the collision integral $C_{12}[f,\Phi]$ describing
collisions of condensate atoms with the thermal atoms, which also
enters the approximate semi-classical kinetic equation for the
single-particle distribution function (valid for $k_{\rm{B}} T \gg
\mu_{c0}$)~\cite{Zaremba99}
\bea 
{\partial f(\bp,\br,t) \over \partial t} &+& {\bp \over m}
\cdot \nabla_\br f(\bp,\br,t)- \nabla U(\br) \cdot
\nabla_{\bp} f(\bp,\br,t) \nonumber \\ &=& C_{12}[f,\Phi] +
C_{22}[f]\,.
\label{eq3}
\eea Here the collision integral denoted by $C_{22}[f]$ describes
binary collisions between non-condensate atoms. It does not change the
number of condensate atoms and hence does {\emph{not}} appear
explicitly in the GP equation \eq{1}. These coupled
equations~(\ref{eq1})-(\ref{eq3}) (along with expressions \eq{a1} and
\eq{a2} in Appendix A for the collision integrals $C_{12}$ and
$C_{22}$) were derived in the semi-classical approximation. They
assume that the atoms in the thermal cloud are well-described by the
single-particle Hartree-Fock spectrum $\tilde \varepsilon_p(\br,t) =
p^2/2m + U(\br,t)$, where $U(\br,t)=U_{\rm ext}(\br) + 2 g [n_c(\br,t)
+ \tilde{n}(\br,t)]$. However, they are expected to contain the
essential physics in trapped Bose-condensed gases at finite $T$, in
both the collisionless and hydrodynamic domains. 

If the gas is weakly disturbed from equilibrium,
one can consider linearized collective oscillations about equilibrium
by writing
\bea
\Phi(\br,t) &=& e^{-i\mu_{c0}t/\hbar}\big[\Phi_0(\br) + \delta 
\Phi(\br,t)\big ], \label{eqnew9} \\
f(\bp, \br, t) &=& f^0(\bp, \br) + \delta f(\bp, \br, t) . 
\label{eqnew10}
\eea Substituting \eq{new9} and \eq{new10} into Eqs.~\eq{1}-\eq{3},
one can obtain the coupled dynamical equations for the damped
oscillations of the condensate $\delta \Phi(\br,t)$ and non-condensate
$\delta f(\bp, \br,t)$. We find that the damping term $R(\br,t)$ in
\eq{2} is finite even if we only keep the equilibrium part of
$f(\bp,\br,t)$. This approximation simplifies the problem tremendously
and should give a good first estimate since $\delta f(\bp,\br,t) \ll
f^0(\bp,\br)$.

The thermal cloud influences the dynamics of the condensate through
two terms in \eq{1}, the mean-field interaction potential
$2g\tilde{n}(\br,t)$ and the second-order collisional term
$iR(\br,t)$.  Let us first consider the effect of the term
$2g\tilde{n}(\br,t)$. In the static approximation, the term $2 g
\tilde n_0(\br)$ gives rise to a temperature-dependent frequency shift
of the condensate collective modes, but does not give rise to
damping. Landau damping appears only when the dynamics of the
non-condensate $\delta \tilde n(\br,t)$ is accounted for (i.e. when we
keep $\delta f(\bp,\br,t)$). An important point is that the
collisional damping described by the second-order term $iR(\br,t)$ in
\eq{1} arises already within the static approximation $\delta
f(\bp,\br,t)=0$, in contrast to Landau damping. 

The damped collective mode equations for the condensate can be 
obtained if we substitute the equilibrium
distribution $f^0(\bp,\br)$ and density $\tilde n_0(\br)$ of the
non-condensate into \eq{1}, giving us
\bea 
i\hbar {\partial \Phi(\br,t) \over \partial t}
&=& \Big [ - {\hbar^2 \over 2 m} \nabla^2 + U_{\rm{ext}}(\br) + g
n_c(\br,t) + 2g \tilde{n}_0(\br) \nonumber \\&-& i R_0(\br,t) \Big
] \Phi(\br,t) , 
\label{eq1b} 
\eea 
which describes the condensate motion coupled to the static thermal
cloud. Here the damping term $R_0(\br,t)$ is 
\beq 
R_0(\br,t) \equiv {\hbar
\over {2 n_c(\br,t)}}\int {d\bp \over (2 \pi \hbar)^3}
C_{12}[f^0(\bp,\br),\Phi(\br,t)].
\label{eq2b} 
\eeq Notice that $R_0(\br,t)$ now depends on time only through
$\Phi(\br,t)$, which has to be determined self-consistently
by solving \eq{1b}.

The equilibrium stationary solution of the coupled equations
\eq{1}-\eq{3} is given by the solution of the generalized
time-independent Gross-Pitaevskii equation \beq \hat{H}_0 \Phi_0(\br)
= \mu_{c0} \Phi_0(\br),
\label{eqnew4}
\eeq 
where we have defined the Hermitian operator 
\beq 
\hat{H}_0 = - {\hbar^2 \over 2 m} \nabla^2 + U_{\rm{ext}}(\br) + 
g n_{c0}(\br) + 2g \tilde{n}_0(\br) .
\label{eqnew5}
\eeq Here, $n_{c0}(\br)=|\Phi_0(\br)|^2$ is the static equilibrium
density of the condensate and $\tilde{n}_0(\br)$ is the equilibrium
density of the non-condensate.  The equilibrium chemical potential of
the condensate $\mu_{c0}$ is independent of position and can be
written explicitly in terms of the equilibrium densities as \beq
\mu_{c0} = -{\hbar^2{\nabla}^2\sqrt{n_{c0}(\br)}\over 
2m\sqrt{n_{c0}(\br)}} + U_{\rm ext}(\br) +gn_{c0}(\br) 
+ 2g\tilde{n}_0(\br) ,
\label{eqnew6}
\eeq where the first term on the right-hand side of \eq{new6} is the
so-called quantum pressure of the static condensate wave function. 

To be consistent with the underlying kinetic model developed in
Ref.~\cite{Zaremba99}, the static thermal cloud is described using the
single-particle HF spectrum (see above). Thus the equilibrium
distribution $f^0(\bp,\br)$ of the non-condensate atoms is given by
the Bose-Einstein distribution \beq f^0(\bp,\br) = {1 \over
{e^{\beta[p^2 /2 m + U_0(\br) - \tilde{\mu}_0]}-1}} \,,
\label{eq4} 
\eeq where $\beta = 1/k_B T$, $\tilde{\mu}_0$ is the equilibrium
chemical potential of the non-condensate, and
$U_0(\br)=U_{\rm{ext}}(\br) +2g[n_{c0}(\br) + \tilde n_0(\br)]$. As
discussed in Ref.~\cite{Zaremba99}, the static chemical potentials of the
two components are equal ($\tilde{\mu}_0 = \mu_{c0}$)
in equilibrium. 

The equilibrium density of the non-condensate $\tilde n_0(\br)$ is
obtained by integrating \eq{4} over momentum to give the usual
result
\beq
\tilde n_0(\br) = {1 \over {\Lambda^3}} g_{3/2}[z_0(\br)].
\label{eqnew8}
\eeq
Here, $\Lambda=(2\pi\hbar^2 \beta/m)^{1/2}$ is the thermal de Broglie
wavelength and $g_{3/2}(z)$ is a Bose-Einstein function.  The local
fugacity is $z_0(\br) = e^{\beta(\tilde \mu_0 - U_0(\br))}$.

We recall that in the TF approximation used in our earlier
work~\cite{Williams2000}, \eq{new6} reduces to $\mu_{c0} =
U_{\rm{ext}} + g n_{c0} +2 g \tilde{n}_0$ and hence $z_0(\br) =
e^{-\beta g n_{c0}(\br)}$. When we keep the quantum pressure terms in
\eq{new6}, the analysis is more complicated. The equilibrium solutions
for the condensate and thermal cloud, given by \eq{new4} and \eq{4},
must be obtained self-consistently.  For completeness, we outline our
procedure for a given value of the temperature $T$ and with the total
number of atoms $N$ fixed (see also Ref.~\cite{Giorgini97}):
\begin{enumerate}
\item For a given value of the condensate population $N_c\leq N$, the
equilibrium solution of the condensate is obtained by solving
\eq{new4} numerically~\cite{Williams99}. To start the procedure, we
initially take $\tilde n_0=0$.
\item Using $\mu_{c0}$ and $n_{c0}(\br)=|\Phi_0(\br)|^2$ found from
step 1, the non-condensate density $\tilde n_0(\br)$ is obtained from
\eq{new8}, using $\tilde \mu_0=\mu_{c0}$. If $\mu_{c0} > {\rm{min}}
\big \{ U(\br) \big \}$, which can occur for small condensates at
temperatures close to $T_{\rm{BEC}}$, we set $\tilde \mu_0 =
{\rm{min}} \big \{ U(\br) \big \}$.
\item Steps 1 and 2 are repeated until convergence is reached to the
desired accuracy (in our calculations, when $\mu_{c0}$ stops varying
up to an error of $10^{-5}$).
\item The number of atoms in the thermal cloud is calculated
$\tilde{N} = \int d\br \tilde n_0(\br)$. We then repeat steps 1 - 3,
varying $N_c$ until the chosen total number of atoms, $N = N_c +
\tilde{N}$, is obtained to the desired accuracy (in our calculations,
we obtain $N$ to an error of less than $10^{-4}$).
\end{enumerate}
The equilibrium values of $n_{c0}(\br)$, $\mu_{c0}=\tilde{\mu}_0$, and
$\tilde n_0(\br)$ obtained from the above procedure
are used in the calculation of the damped collective oscillations of
the condensate discussed in the next section. 

Before proceeding, though, it is useful to emphasize several points
about our approximate model, as described by \eq{1b} and \eq{2b}. Our
emphasis in this paper, and the earlier TF version in
Ref.~\cite{Williams2000}, is on the calculation of damping due to
collisional exchange of atoms between a dynamic condensate and a
static thermal cloud. While we use the semi-classical HF excitations
to calculate $f^0(\bp,\br)$, which leads to \eq{new6} and \eq{4}, we
do not expect improved treatments of the static thermal cloud will
greatly alter our estimates of the inter-component condensate damping
we are considering. There is a considerable literature on the
calculation of the undamped condensate mode frequencies at finite
$T$. One finds that the frequency shifts are quite dependent on the
specific approximation~\cite{Hutchinson97,Dodd98,Dodd98b} used for the
non-condensate mean field $2 g \tilde{n}_0(\br)$, as well as on other
terms left out of \eq{1b} related to the anomalous correlation
function $\tilde m_0(\br) = \langle \tilde \psi(\br) \tilde
\psi(\br) \rangle$ We are not concerned with these questions here, but
refer to Refs.~\cite{Giorgini2000,Bergeman2000} for further studies.

\section{Bogoliubov equations with damping}

Using the explicit general expression for $C_{12}$
given in \eq{a2} of Appendix A, it was shown in Ref.~\cite{Zaremba99}
that $R_0(\br,t)$ can be simplified to (see also
Ref.~\cite{Gardiner2000}) \beq R_0(\br,t) = {\hbar
\over{2\tau_{12}(\br,t)}} \big [e^{-\beta (\tilde{\mu}_0
-\varepsilon_c(\br,t))} - 1 \big ],
\label{eqx1}
\eeq 
where we have defined the relaxation rate due to inter-component
$C_{12}$ collisions
\bea 
{1 \over {\tau_{12}(\br,t)}} &\equiv& {2 g^2 \over
(2\pi)^5\hbar^7} \!\int \!\!d{\bf p}_1 \!\int \!\!d{\bf p}_2 \!\int
\!\!d{\bf p}_3 \delta(\bp_c+{\bf p}_1-{\bf p}_2-{\bf p}_3) \nonumber
\\ &\times& \delta(\varepsilon_c+\tilde \varepsilon_{p_1} -\tilde
\varepsilon_{p_2}-\tilde \varepsilon_{p_3}) (1+f_1^0) f_2^0 f_3^0 \,.
\label{eqx2}
\eea 
The condensate atom local energy is $\varepsilon_c(\br,t) =
\mu_c(\br,t) + {1\over 2}m v_c^2(\br,t)$ with the non-equilibrium
condensate chemical potential 
\beq 
\mu_c(\br,t) =
-{\hbar^2{\nabla}^2\sqrt{n_c(\br,t)}\over 2m\sqrt{n_c(\br,t)}} + 
U_{\rm ext}(\br) +gn_c(\br,t) + 2g\tilde{n}_0(\br)\,.
\label{eqx3}
\eeq 
The condensate atom momentum is $\bp_c = m \bv_c$, and $f_i^0 =
f^0(\br,\bp_i)$. We have introduced the usual condensate velocity
defined in terms of the phase $\theta$ of the condensate $\Phi(\br,t)
= \sqrt{n_c(\br,t)} e^{i \theta(\br,t)}$ as ${\bf v}_c =
\hbar{\nabla}\theta(\br,t)/m$. A closed set of equations for
$\Phi(\br,t)$ is given by \eq{1b} combined with \eq{4} and \eq{x1} -
\eq{x3}. The damping term $R_0(\br,t)$
given in \eq{x1} vanishes when the condensate is in diffusive 
equilibrium with the thermal cloud, when 
$\varepsilon_c(\br,t) = \mu_{c0} = \tilde{\mu}_0$.

We obtain the linearized equation of motion for the
condensate fluctuation $\delta \Phi(\br,t)$ by expanding  all condensate
variables appearing in \eq{1b} to first order in $\delta \Phi(\br,t)$.
For simplicity we assume $\Phi_0(\br)$ is real, 
that is, $\bv_{c0}(\br)=0$. Using \eq{new9}, the condensate
density $n_c(\br,t) = |\Phi_0(\br) + \delta \Phi(\br,t)|^2$ can be
written $n_c(\br,t) = n_{c0}(\br) + \delta n_c(\br,t)$,
where the linearized density fluctuation of the condensate is
\beq
\delta n_c(\br,t) = \Phi_0(\br) \big [\delta \Phi(\br,t) + 
\delta \Phi^*(\br) \big ].
\label{eqdensfluct}
\eeq We can also simplify $R_0(\br,t)$ in \eq{x1} by noting that, to
first order in the condensate fluctuations, we can write
$\varepsilon_c(\br,t) = \mu_c(\br,t)$ (neglecting the quadratic term
$(\delta v_c)^2$). Using \eq{new6}, the
condensate chemical potential \eq{x3} can be written as 
\beq \mu_c(\br,t) = \mu_{c0} + \delta \mu_c(\br,t),
\label{eqmuexpand}
\eeq
where the linearized fluctuation in the local 
condensate chemical potential is found to be given by
\beq 
\delta \mu_c(\br,t) = {1\over{2\Phi_0}}(\hat{H}_0 - 
\mu_{c0})\Big [{\delta n_c(\br,t)\over\Phi_0(\br)}
\Big ] + g \delta n_c(\br,t). 
\label{eq13} 
\eeq where the operator $\hat{H}_0$ is defined in \eq{new5}. The first
term in \eq{13} arises from the dynamic quantum pressure of the
condensate oscillation, while the second term comes from the mean
field interaction. The latter is the only contribution kept in the TF
approximation used in Ref.~\cite{Williams2000}. Note that because the
non-condensate is static, there is no contribution in \eq{13} from the
fluctuation in the HF mean field ($2g\,\delta \tilde n$) of the
thermal cloud.

Utilizing the fact that $\mu_{c0}=\tilde \mu_0$ in static 
equilibrium, $R_0(\br,t)$ in \eq{x1} can be simplified to 
\beq \delta
R_0(\br,t) = {\hbar \beta \over {2 \tau_{12}^0(\br)}} \delta \mu_c(\br,t) \,,
\label{eq11}
\eeq 
where the ``equilibrium'' $C_{12}$ collision rate is defined by 
(compare with \eq{x2})
\bea 
{1\over{\tau^0_{12}(\br)}} &\equiv& {2 g^2 \over 
(2\pi)^5\hbar^7} \int d{\bf p}_1 \int d{\bf p}_2 
\int d{\bf p}_3 \delta({\bf p}_1-{\bf p}_2-{\bf
p}_3) \nonumber \\ &\times&\delta \big({p_1^2-p_2^2-p_3^2\over{2m}}-g
n_{c0}\big ) (1+f_1^0) f_2^0 f_3^0 \,.
\label{eq12} 
\eea 

We can now obtain an equation of motion for the condensate fluctuation
$\delta \Phi(\br,t)$ by substituting \eq{new9} into \eq{1b} and using
the above results
\bea
i \hbar {\partial \delta \Phi(\br,t) \over {\partial t}} &=& 
\hat{\cal L} \, \delta \Phi(\br,t) + g n_{c0}(\br) \delta \Phi^*(\br,t)
\nonumber \\
&-& i \hat{\Gamma} \, \big [\delta \Phi(\br,t) + \delta \Phi^*(\br,t) \big ],
\label{eqflucteom}
\eea
We find it convenient to define the operators
$\hat{\cal L} = \hat{H}_0 + g n_{c0} - \mu_{c0}$ and  
\beq 
\hat{\Gamma} = {\hbar \beta \over {2 \tau_{12}^0}} \Big[ \frac{1}{2} \big 
(\hat{H}_0 - \mu_{c0} \big ) + g n_{c0} \Big ],
\label{eq16} 
\eeq 
both of which are Hermitian.  The first term in \eq{16} arises from
the quantum pressure of the condensate fluctuation. It is
neglected in the Thomas-Fermi approximation, valid in the large $N_c$
limit. The damping operator can also be written in an alternative form
as $\hat{\Gamma} = (\hbar\beta/ 4 \tau_{12}^0)[\hat{\cal L} + g
n_{c0}(\br)]$.

We now consider normal mode oscillations of the condensate by 
expanding $\delta \Phi(\br,t)$ as
\beq 
\delta \Phi(\br,t) =
u_j(\br) e^{-i\varepsilon_jt/\hbar} + 
v_j^*(\br) e^{i\varepsilon_j^*t/\hbar} ,
\label{eq10}
\eeq where the excitation energy $\varepsilon_j$ and the collective
mode amplitudes $u_j(\br)$ and $v_j(\br)$, in general, have real and
imaginary components. Substituting this into \eq{flucteom} and
equating like powers of $e^{\pm i \varepsilon_j t/\hbar}$, we obtain
the damped finite-$T$ Bogoliubov equations \bea \big ( \hat{\cal L} -
i \hat{\Gamma} \big ) u_j(\br) + \big (g n_{c0} - i \hat{\Gamma} \big
) v_j(\br) &=& \varepsilon_j u_j(\br) \nonumber \\ \big (\hat{\cal L}
+ i \hat{\Gamma} \big ) v_j(\br) + \big (g n_{c0} + i \hat{\Gamma}
\big ) u_j(\br) &=&-\varepsilon_j v_j(\br) .
\label{eq15}
\eea The real and imaginary parts of $\varepsilon_j$ give the
oscillation frequency $\omega_j \equiv {\rm{Re}} \varepsilon_j/\hbar $
and the damping rate $\gamma_j \equiv -{\rm{Im}} \varepsilon_j/\hbar
$, which vary with temperature $T$. These coupled equations are the
main new result of this paper. They describe the damped collective
modes of a condensate coupled to a static thermal cloud at finite
$T$. It is important to realize that the anti-Hermitian operator $i
\hat{\Gamma}$ appearing in \eq{15} has not simply been inserted ``by
hand'' based on phenomenological arguments, as is done in the theories
developed in Refs.~\cite{Pitaevskii59,Hutchinson99,Choi98}, but has
been derived explicitly starting from a microscopic (albeit
approximate) theory. The origin of this damping is the lack of
collisional detailed balance between the condensate and
non-condensate, which occurs when the system is disturbed from
equilibrium. This mechanism is distinct from the Landau damping
process, which arises from the mean field interaction between
components when the dynamics $\delta f(\bp,\br,t)$ of the
non-condensate is taken into account~\cite{Giorgini2000}.

In order to explore the properties of the coupled Bogoliubov equations
\eq{15}, it is useful to write it in matrix form \beq
\hat{M} w_j = \varepsilon_j \sigma_3 w_j, \label{eq17}\eeq where
$\hat{M} = \hat{M}_0 + \hat{M}'$ and
\begin{equation}
\begin{array}{ccc}
\hat{M}_0 &=&
\left( \begin{array}{cc}  
	\hat{\cal L} & g n_{c0}  \\ 
	g n_{c0}   & \hat{\cal L} 
\end{array} \right) ,
\end{array}
\end{equation}
\begin{equation}
\begin{array}{ccc}
\hat{M}' &=& -i \hat{\Gamma}
\left( \begin{array}{cc}  
	1 & 1  \\ 
	-1&  -1
\end{array} \right) ,
\end{array}
\label{eq19}
\end{equation}
\begin{equation}
\begin{array}{ccc}
\sigma_3 &=&
\left( \begin{array}{cc}  
	1  & 0 \\ 
	0 & -1
\end{array} \right) ,
\end{array}
\end{equation}
\begin{equation}
\begin{array}{ccc}
w_j(\br) &=&
\left( \begin{array}{c}  
	u_j(\br)  \\ 
	v_j(\br)
\end{array} \right) .
\end{array}
\end{equation}
Since $\hat{M}'$ is not a Hermitian operator, the solutions of \eq{17}
are not in general orthogonal to one another and 
the energies $\varepsilon_j$ are expected to have an imaginary
component describing damping. We can see this explicitly by deriving a
generalized orthogonality relation for the solutions 
$w_j$~\cite{Fetter72,Blaizot}.
Multiplying \eq{17} on the left-hand side by
$w_i^\dagger$ and integrating over position, we find
\beq
\int d\br w_i^\dagger(\br) \hat{M} w_j(\br) = 
\varepsilon_j \int d\br w_i^\dagger(\br) \sigma_3 w_j(\br).
\label{eqorth1}
\eeq
Taking the Hermitian conjugate of \eq{17}, replacing the
index $j$ by $i$, multiplying on the right-hand side by $w_j$,
and integrating over position, we obtain
\beq
\int d\br w_i^\dagger(\br) \hat{M}^\dagger w_j(\br) = 
\varepsilon_i^* \int d\br w_i^\dagger(\br) \sigma_3 w_j(\br).
\label{eqorth2}
\eeq
Subtracting \eq{orth1} from \eq{orth2} gives us
the desired orthogonality relation
\beq 
(\varepsilon_{i}^* - \varepsilon_{j})\int d\br
w_{i}^\dagger(\br) \sigma_3 w_{j}(\br) = 2i \int d\br w_{i}^\dagger
(\br)\sigma_3 \hat{\Gamma} w_{j}(\br). 
\label{eq25} 
\eeq
where we have used the Hermitian property $\hat{M}_0 = \hat{M}_0^\dagger$
and used the fact that
$(\hat{M}'^\dagger - \hat{M}')=2i \sigma_3 \hat{\Gamma}$.

If we neglect the damping term $\hat{\Gamma} = 0$ in \eq{17}, our
description reduces to the usual coupled Bogoliubov equations at
finite temperatures for the collective modes $u_{0j}(\br)$ and
$v_{0j}(\br)$ in the static Popov
approximation~\cite{Hutchinson97,Dodd98,Dodd98b,Griffin96} \beq
\hat{M}_0 w_{0j} = \varepsilon_{0j} \sigma_3 w_{0j}.
\label{eq22}
\eeq 
With $\hat{\Gamma}=0$ in \eq{25}, we see that these solutions
$w_{0j}$ obey the orthogonality condition
\beq 
(\varepsilon_{0i}^* - \varepsilon_{0j})\int d\br
w_{0i}^\dagger \sigma_3 w_{0j} = 0.
\label{eq23} 
\eeq It then follows that the eigenvalues must be real
($\varepsilon_{0j} = \varepsilon_{0j}^*$), otherwise the solution
$w_{0j}$ of \eq{22} would have zero norm~\footnote{Complex frequencies
of the usual Bogoliubov equations (with no explicit imaginary term)
can arise if one is considering collective oscillations of a
dynamically unstable order parameter, such as a multi-component
condensate~\cite{Law97,Garcia-Ripoll2000} or one involving
solitons~\cite{Feder2000}.}. This property follows from the fact that
$\hat{M}_0$ is a Hermitian operator.  We note that at $T=0$, where
$2g\tilde{n}_0=0$ in $\hat{M}_0$, \eq{22} reduces to the standard
Bogoliubov equations~\cite{Fetter72,Blaizot}. At finite temperatures,
the simplest way of including the non-condensate is to use the
temperature dependent condensate number $N_c(T)$ (but ignore the
mean-field term $2g\tilde{n}_0(\br)$ in
$\hat{H}_0$)~\cite{Dodd98b}. As discussed in \cite{Williams2000}, the
condensate normal mode frequencies at finite $T$ are the same as at
$T=0$ in the Thomas-Fermi limit~\cite{Stringari96}, since the
frequencies at $T=0$ do not depend on the value of $N_c$ in this
limit. Another effect comes from the mean field $2 g \tilde{n}_{0}$
due to the non-condensate, which gives rise to a temperature-dependent
energy shift to the excitation
frequencies~\cite{Hutchinson97,Dodd98}. However, as discussed at the
end of Section II, a proper estimate of the shifts in the condensate
normal mode frequencies requires a better model.

According to \eq{25}, the inclusion of the anti-Hermitian damping
operator $i \hat{\Gamma}$ in the normal
mode equations introduces an explicit imaginary part in the
eigenenergies $\varepsilon_j$, without requiring the solutions to have
zero norm. However, these damped solutions $w_i$ 
are not required to be orthogonal. From \eq{25}, the imaginary part 
of $\varepsilon_j$ is given by 
\beq
{\rm{Im}}  \varepsilon_j 
 = {{- \int d\br
w_j^\dagger \sigma_3 \hat{\Gamma} w_j }\over {\int d\br w_j^\dagger
\sigma_3 w_j}} .  
\label{eq26} 
\eeq If the operator $\hat{\Gamma}$ in \eq{16} was a scalar
position-independent constant $\hat{\Gamma} = \gamma$, we can take
it out of the integral in \eq{25}, and in this case the orthogonality
condition for $w_j$ is recovered, \beq (\varepsilon_{i}^* -
\varepsilon_{j}-2i \gamma)\int d\br w_{i}^\dagger \sigma_3 w_{j} = 0.
\label{eq27} 
\eeq This states that for $i=j$, the imaginary part of each of the
excitation energies is ${\rm{Im}} \varepsilon_j = - \gamma$ (unless
the norm of $w_j$ is zero). For $i\neq j$, \eq{27} shows that the
eigenmodes $w_i$ and $w_j$ are orthogonal as long as ${\rm{Re}}
\varepsilon_i \neq {\rm{Re}} \varepsilon_j $.

Since the damping effect of $C_{12}$ collisions on the condensate
oscillations is weak, we can treat the damping term $\hat{M}'$ in
\eq{19} using first order perturbation theory. More precisely, we
assume \beq \hbar \omega_j \gg \gamma_j,
\label{eqclessnew}
\eeq
where $\gamma_j = -{\rm{Im}}  \varepsilon_j $ is the
damping rate due to $C_{12}$ collisions given by \eq{26}. 
Proceeding as usual in perturbation theory, we introduce
a small expansion parameter
\beq
\hat{M} = \hat{M}_0 + \lambda \hat{M}',
\label{eqexpM}
\eeq
and expand the energies and eigenmodes as
\bea
\varepsilon_j &=& \varepsilon_{0j} + \lambda \varepsilon_{1j}
 + \lambda^2 \varepsilon_{2j} + \cdots \label{eqexp1} \\
w_i &=& w_{0j} + \lambda w_{1j} + \lambda^2 w_{2j} +
\cdots.
\label{eqexp2}
\eea Substituting \eq{expM} - \eq{exp2} into \eq{17} and equating like
powers of $\lambda$, we obtain a hierarchy of equations.  Note that
since $\hat{M}'$ is also radially symmetric, it does not break the
$(2l+1)$-fold degeneracy of the magnetic sublevels. The zeroth order
equation simply reproduces \eq{22} for the undamped modes
$w_{0j}$. The first order correction is obtained from \beq \big
(\hat{M}_0 - \varepsilon_{0j} \sigma_3 \big ) w_{1j} = \big (
\varepsilon_{1j} \sigma_3 - \hat{M}' \big ) w_{0j} .
\label{eqpoop}
\eeq
If we multiply \eq{poop} on the left by $w_{0j}^\dagger$ and
integrate over position, the left-hand side of \eq{poop}
vanishes using \eq{22}. This leaves us with the
following result for the perturbed energy shift,
\beq
\varepsilon_{1j} = {\int d \br w_{0j}^\dagger \hat{M}' w_{0j} 
\over {\int d \br w_{0j}^\dagger \sigma_3 w_{0j}}}.
\label{eqcorr1}
\eeq Taking the Hermitian conjugate of \eq{poop}, multiplying on the
right by $w_{0j}$ and integrating over position, we obtain the
analogous expression for the complex conjugate $\varepsilon_{1j}^*$
\newpage
\beq \varepsilon_{1j}^* = {\int d \br w_{0j}^\dagger \hat{M}'^\dagger
w_{0j} \over {\int d \br w_{0j}^\dagger \sigma_3 w_{0j}}}.
\label{eqcorr2}
\eeq

From \eq{corr1} and \eq{corr2}, we can obtain the damping
(to first order)
\beq 
 \gamma_{0j} = -{\rm{Im}} \varepsilon_{1j}
= {\int d\br w_{0j}^\dagger \sigma_3 \hat{\Gamma} w_{0j}
\over {\int d \br w_{0j}^\dagger \sigma_3 w_{0j}}} .
\label{eq28} 
\eeq 
Here we have used $(\hat{M}'^\dagger - \hat{M}')=
2i \sigma_3 \hat{\Gamma}$. The result in \eq{28} 
is consistent with expression
\eq{26}, had we simply expanded \eq{26} about $w_{0j}$. We can
also obtain the real part of $\varepsilon_{1j}$ from \eq{corr1}
and \eq{corr2} and we find that this vanishes (see Appendix B), i.e.
\beq
{\rm{Re}}  \varepsilon_{1j} = 0.
\label{eqrealzero}
\eeq
In summary, first order perturbation theory gives
\beq
\varepsilon_j = \varepsilon_{0j} - i \gamma_{0j} .
\eeq
Using these results, the expansion given in 
\eq{10} for the condensate oscillations can be written
in the more explicit form
\beq 
\delta \Phi(\br,t) = e^{-\gamma_{0j}t/\hbar} \big (
u_{0j}(\br) e^{-i\varepsilon_{0j} t/\hbar} + 
v_{0j}^*(\br) e^{i\varepsilon_{0j} t/\hbar} \big ) .
\label{eq10b}
\eeq 
Using expression \eq{28} to calculate $\gamma_{0j}$, one can
check that the criterion in \eq{clessnew} is well satisfied. The
difference between the perturbative results and the exact numerical
solution of \eq{15} is always less than $1 \, \%$ in our calculations.

The results of the present calculation can be compared to those in
Ref.~\cite{Williams2000}, where we used a simplified model to
calculate the damping rates $\gamma_j$.  We used the Thomas-Fermi
approximation (neglecting the kinetic energy pressure in the solution
of $\Phi_0$ and $\mu_{c0}$) and we also neglected the contribution $2
g \tilde n_0$ to the mean field interaction. We also neglected the
kinetic energy pressure in the solution of the normal modes, which
allowed us to use the $T=0$ Stringari normal mode
solutions~\cite{Stringari96} as a basis with which to treat the effect
of $\delta R_0$ to first order.  The main result of
Ref.~\cite{Williams2000} was the expression for the damping
rate \beq \gamma_j = {\hbar \over 2} {\int d \br \, \delta
n_{{\rm{S}}j}^2(\br) /\tau'(\br)\over \int d\br \, \delta
n_{{\rm{S}}j}^2(\br)}.
\label{eq29} 
\eeq Here $\delta n_{{\rm{S}}j}(\br)$ is the density fluctuation
associated with a Stringari normal
mode~\cite{Stringari96,Williams2000} at finite $T$ and $1/\tau' = (g
n_{c0}/ k_B T) (1/ \tau_{12}^0)$, with $1/\tau_{12}^0$ given in
\eq{12}. This damping rate is easy to evaluate since the equilibrium
form for the condensate $\Phi_0(\br)$, as well as the Stringari normal
modes $\delta n_{\rm{S}j}(\br)$, are both given by simple analytic
functions.

The TF result \eq{29} can, of course,
be obtained directly from \eq{28} by taking
the Thomas-Fermi limit. In this limit, the first term in \eq{16} for
$\hat{\Gamma}$, which describes
 the quantum pressure of the oscillation, can be
neglected. We are left with $\Gamma(\br) = \hbar/2\tau'(\br)$,
and hence \eq{28} reduces to 
\beq 
\gamma_{0j} = {\hbar \over 2} {\int
d\br w_{0j}^\dagger \sigma_3 w_{0j}/\tau'(\br) \over {\int d \br
w_{0j}^\dagger \sigma_3 w_{0}}}.  
\label{eq53}
\eeq 
In Ref.~\cite{Fetter98}, the
product $w_{0j}^\dagger(\br) \sigma_3 w_{0j}(\br) = |u_j(\br)|^2 -
|v_j(\br)|^2$ is shown to satisfy the following relationship \beq
|u_j(\br)|^2 - |v_j(\br)|^2 = {\rm{Re}}  {-2 i m \over \hbar}
\delta\phi_j^* (\br)\delta n_j(\br) ,
\label{eq54}
\eeq where $\delta n_j(\br)$ is the density fluctuation of mode $j$,
and $\delta\phi_j(\br)$ is the corresponding fluctuation in the
velocity potential. In the Thomas-Fermi limit, the finite $T$
fluctuations in the density and velocity potential still have the simple
relationship~\cite{Wu96} \beq \delta \phi_{{\rm{S}}j}(\br) = {-i g
\over {m \omega_{{\rm{S}}j}}} \delta n_{{\rm{S}}j}(\br) ,
\label{eq55}
\eeq
where $\omega_{{\rm{S}}j}$ is the Stringari frequency 
of mode $j$. Putting these results together, we have
\beq
w_{0j}^\dagger(\br) \sigma_3 w_{0j}(\br) = {2 g \over {\hbar \omega_j}}
\delta n_{{\rm{S}}j}^2(\br)
\label{eq56}
\eeq and hence \eq{53} reduces to \eq{29}. 

In the next section, we compare the damping given by \eq{29} to the
result given by \eq{28} for a trapped gas. It is useful to apply
\eq{28} to the case of a homogeneous gas, where the modes are
$u_\bk(\br) = u_k e^{i \bk \cdot \br}$ and $v_\bk(\br) = v_k e^{i \bk
\cdot \br}$ and the energies are given by the usual Bogoliubov
expression \beq \varepsilon_k = \sqrt{\Big({\hbar^2 k^2 \over {2m}}
\Big ) \Big ({\hbar^2 k^2 \over{2 m}} + 2g n_{c0} \Big )} .  \eeq
Using the expression for $\hat{\Gamma}$ given in \eq{16}, the damping
rate \eq{28} reduces to \beq \gamma_k = {\hbar \beta \over {2
\tau_{12}^0}} \Big[ {\hbar^2 k^2 \over {4 m}}+ g n_{c0} \Big ], \eeq
where $\tau_{12}^0$, as given in \eq{12}, is independent of position
for a uniform Bose gas. The first term proportional to $k^2$ comes
from the quantum pressure (i.e. the first term in \eq{16}), which is
omitted in the Thomas-Fermi result given in Ref.~\cite{Williams2000}.

\newpage
\section{Numerical Results}\label{results}

We now turn to an explicit calculation of the inter-component damping
of collective modes for a dilute Bose gas in a spherical harmonic trap
$U_{\rm{ext}}(\br) = \frac{1}{2}m\omega_0^2 r^2$. We choose the same
set of physical parameters as given in Ref.~\cite{Williams2000}: the
frequency of the trap is $\omega_0/{2\pi} = 10$ Hz, the scattering
length for ${}^{87}$Rb is $a = 5.7$ nm, and the total number of atoms
is $N=2 \times 10^6$.

\subsection{Equilibrium solution}
In Fig.~1 we plot the condensate fraction versus temperature for the
case of $N = 2 \times 10^6$ atoms.  The solid line corresponds to the
full calculation of $N_c(T)$ as described in steps 1-4 of Section
II. The dashed line is obtained using the Thomas-Fermi approximation
for $\Phi_0$ and neglecting the effect of $2g\tilde n_0$ in
$\hat{H}_0$ and $U(\br)$, as described in
Refs.~\cite{Minguzzi97a,Williams2000}.  The two curves deviate only
slightly as $T$ approaches $T_{\rm{BEC}}$. This is consistent with
previous studies~\cite{Giorgini97,Holzmann99}, which found that the
thermodynamic quantities of a dilute Bose gas are quite insensitive to
the level of approximation used to treat the non-condensate
spectrum. In Fig. 1, for comparison, we also show the result for the
case of the ideal gas in the thermodynamic limit, $N_c/N = [1 -
(k_{\rm{B}} T / \hbar \omega_0)^3]$~\cite{Dalfovo99}, given by the
dot-dashed line. When interactions are included, the effective
condensation temperature $T_{\rm{BEC}}$ is slightly
lower~\cite{Giorgini97}.

\begin{figure}
  \centerline{\epsfig{file=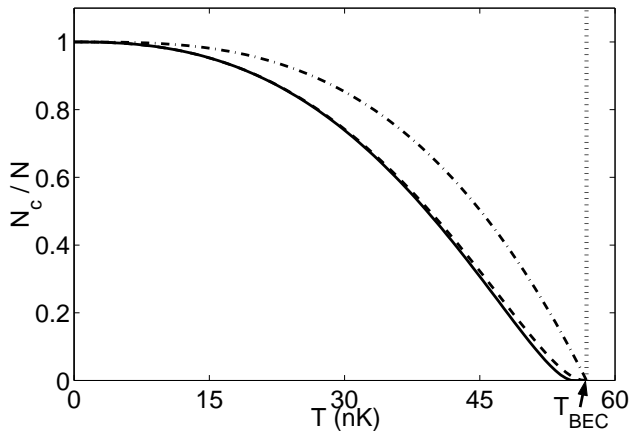,width=3.3in}}
\caption{Condensate fraction versus temperature for $N=2 \times 10^6$
atoms. The solid line corresponds to the method described in steps 1-4
in {Section~II}. The Thomas-Fermi approximation is used to calculate
the dashed line, as described in
Ref.~\cite{Williams2000,Minguzzi97a}. The dot-dashed line is for an
ideal Bose gas in the thermodynamic limit. }
\end{figure}

\subsection{Collective mode frequencies}
\begin{figure}
  \centerline{\epsfig{file=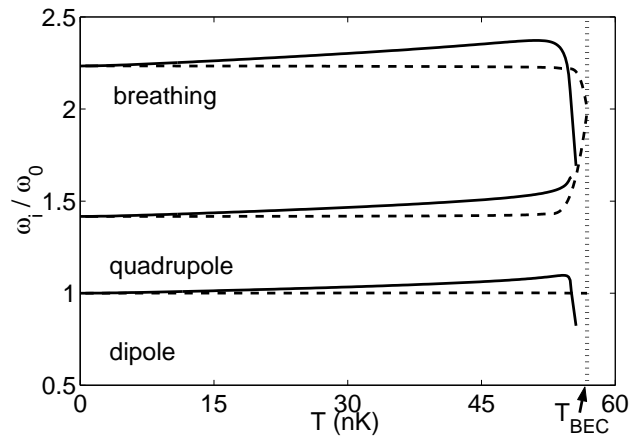,width=3.3in}}
\caption{Condensate mode frequencies versus temperature for $N=2
\times 10^6$ atoms. The dashed lines are the $T=0$ modes but
for a changing number of condensate atoms $N_c(T)$. The solid
line is the real part of the
eigenenergies $\varepsilon_i$ obtained by solving \eq{17}. }
\end{figure}
We plot the collective excitation frequencies $\omega_i$ in Fig.~2 for
the breathing $\{n=1,l=0,m=0\}$, dipole $\{0,1,m\}$, and quadrupole
$\{0,2,m\}$ modes. The solid line corresponds to the numerical
solution of the full coupled Bogoliubov equations in \eq{17}. For
comparison, the dashed lines are obtained from the solution of \eq{22}
setting $2g\tilde{n}_0=0$ and using the value of $N_c(T)$ (as given by
the dashed line in Fig.~1).  In the Thomas-Fermi Stringari limit, the
frequencies are given by the $T=0$ results, at all temperatures.

As $T$ approaches $T_{\rm{BEC}}$, $N_c(T)$ becomes very small and the
solutions given by the dashed lines go over to the harmonic oscillator
eigenstates---the breathing and quadrupole modes become degenerate at
$\omega_i/\omega_0 = 2$. As $T$ decreases, the number of atoms in the
condensate increases, so that the dashed line approaches the $T=0$
Stringari frequencies.  The dipole mode frequency differs from the
value $\omega_i =\omega_0$ expected for the Kohn mode since the
thermal cloud is treated statically (see also
Ref.~\cite{Hutchinson97}).

The deviation of the full solution (given by the solid lines) from the
dashed lines is due entirely to the mean field of the static thermal
cloud $2g\tilde n_0$. The upward shift in the frequencies (as
illustrated in Fig. 2 for $N=2 \times 10^6$) increases with $N$.
Early studies~\cite{Hutchinson97,Dodd98,Dodd98b} of collective modes
in the static Popov approximation considered much smaller systems, $N
\sim 10^3$.  In more recent work~\cite{Bergeman2000}, the mode
frequencies were computed for a much larger system of $N = 2 \times
10^5$ atoms.  Although the thermodynamic quantities are very
insensitive to the level of approximation used to treat the spectrum
of the non-condensate atoms~\cite{Giorgini97, Holzmann99}, the
lowest-lying collective mode frequencies are clearly more
sensitive. While we feel it is useful to show what our simple model
gives for the frequency shifts, the shifts found in
Refs.\cite{Hutchinson97} and~\cite{Dodd98}, based on calculating
$\tilde n_0(\br)$ more self-consistently, are much smaller than our
model predicts. A good estimate clearly requires a more realistic
theory~\cite{Giorgini2000,Bergeman2000}, as discussed at the end of
Section II.

\subsection{Damping due to inter-component collisions}
In Fig.~3, we plot the damping rates $\gamma_i$ for $N = 2\times 10^6$
atoms.  In the main graph the damping of the breathing mode is shown,
where the solid line comes from a full calculation based on \eq{17}
and the dashed line is calculated using the simplified expression in
\eq{29} derived in the Thomas-Fermi limit~\cite{Williams2000}. As a
reference point, we have indicated the temperature at which the
condensate population $N_c(T)$ reaches $10^4$ atoms, where the ratio
in \eq{TFA} is $r_{\rm{HO}}/R_{\rm{TF}} = 0.33$. The largest
deviation, as expected, occurs close to $T_{\rm{BEC}}$ where the
Thomas-Fermi approximation \eq{TFA} used in deriving \eq{29} starts to
break down. The damping becomes very large as $N_c$ goes to zero close
to the condensation temperature. This increase arises from the kinetic
quantum pressure of the collective oscillation, i.e. the first term in
\eq{16}.

In the inset of Fig.~3 we graph the damping rates for all three modes
considered in Fig.~2. The similarity emphasizes the fact that the
damping is fairly insensitive to the detailed form of the condensate
normal modes, which suggests that the inter-component damping shown in
Fig. 3 will not be significantly modified by a more accurate
description of the {\emph static} thermal cloud, in contrast to the
frequency shifts.
\begin{figure}
  \centerline{\epsfig{file=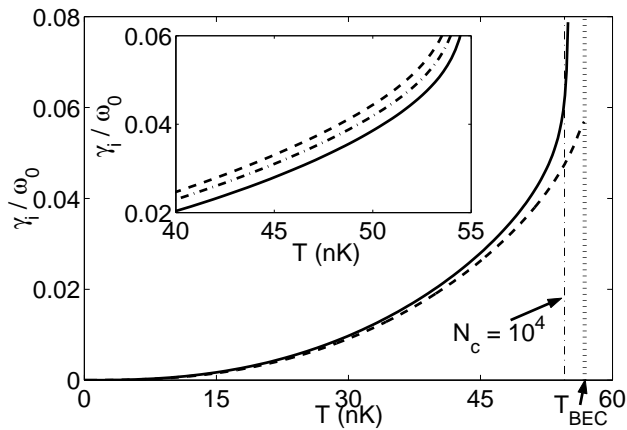,width=3.3in}}
\caption{Damping rates versus temperature for $N=2 \times 10^6$
atoms. In the main plot, the solid line corresponds to the full
calculation of $\gamma_i = -{\rm{Im}} \varepsilon_i$ for the
breathing mode as obtained by solving \eq{17}. The simplified TF
expression \eq{29} is given by 
the dashed line. In the inset, the full calculation of
$\gamma_i$ is shown for the breathing (solid), dipole (dot-dashed),
and quadrupole (dashed) modes.}
\end{figure}

Clearly as we approach $T_{\rm{BEC}}$, the condensate is becoming
small while the thermal cloud is becoming more dominant. In this
transition region, the nature of the excitations in a trapped Bose gas
is quite complex. Since the condensate is disappearing, so are the
collective modes and a theory based on \eq{1} becomes inadequate.

\section{Discussion and Outlook}

In summary, we have derived damped collective mode equations \eq{15}
for a condensate interacting with a static thermal cloud starting from
the generalized GP equation derived in Ref.~\cite{Zaremba99}. This
mechanism is due to the lack of collisional detailed balance between
the two components when the condensate is disturbed from
equilibrium. In a previous article~\cite{Williams2000}, we evaluated
this damping in the large $N_c$ Thomas-Fermi limit and neglected the
mean field of the non-condensate $2g\tilde n_0$. In Section IV we
presented results of an explicit calculation of \eq{15} and verified
that the simplified TF expression \eq{29} agrees quantitatively with
the damping obtained from the full calculation, except very near
$T_{\rm{BEC}}$ where $N_c(T)$ is becoming small.

Starting from the finite $T$ generalized GP equation \eq{1b}, we
formulated our calculation in terms of coupled Bogoliubov equations
\eq{15}, in which the inter-component collision damping arises
explicitly. We have used this opportunity to discuss some of the
formal properties of these equations when damping is present. Such
equations may be of interest in other problems when dealing with
damping from the thermal cloud.

The effect of damping in the time-dependent Gross-Pitaevskii equation
has been discussed in several different
contexts~\cite{Pitaevskii59,Hutchinson99,Choi98}. Many years ago,
Pitaevskii~\cite{Pitaevskii59} developed a phenomenological model for
superfluid helium to describe the evolution of the superfluid toward
equilibrium in the two-fluid hydrodynamic regime.  This model has
points of contact with the work in Ref.~\cite{Zaremba99}.  In
Ref.~\cite{Choi98}, a similar scheme was used to discuss damping of
condensate modes in the collisionless region.  In
Ref.~\cite{Hutchinson99}, damping terms were introduced into the GP
equation to account for output coupling of condensate atoms from the
trap as well as the exchange of atoms between the condensate and
thermal cloud. Finally, in quite a different context, damped
collective modes were calculated in Ref.~\cite{Horak2000} for a
condensate in an optical cavity.

Taking into account the dynamic mean field coupling between
components, the collective modes of the system become a hybridization
of condensate oscillations and non-condensate oscillations, so that
one gets essentially a pair of in-phase and out-of-phase oscillations
for each collective mode. For a given mode symmetry, the out-of-phase
mode consists mostly of condensate oscillations with the collective
motion of the non-condensate being less
significant~\cite{Zaremba99}. Our model calculations (and those of
Ref.~\cite{Williams2000}) should be quite good for the intercomponent
$C_{12}$ damping of such out-of-phase modes. In the other extreme, the
in-phase Kohn mode~\cite{Zaremba99} involves {\emph{both}} the
condensate and non-condensate moving together in local equilibrium, in
which case $R(\br,t)=0$.

The dynamic coupling between the condensate and thermal cloud gives
rise to damping of the condensate oscillations (known as Beliaev and
Landau damping), which are quite distinct mechanisms from that
considered in the present paper. Beliaev and Landau damping arise from
dynamic mean-field effects, rather than collision terms such as
$C_{12}$ or $C_{22}$ (for further discussion and references, see
Refs.~\cite{Giorgini2000,Williams2000}). In a complete theory the
condensate collective mode damping would be given by
$\gamma_{\rm{tot}} = \gamma_{L} + \gamma_{B} + \gamma_{C}$, that is,
Landau, Beliaev, and collisional (due to $C_{12}$) damping,
respectively. Our model gives a good estimate of the $C_{12}$ damping
$\gamma_C$. In our earlier work~\cite{Williams2000} where we used the
TFA, we compared the $C_{12}$ damping directly with Landau damping and
found that $\gamma_{C}/\gamma_{L} \sim 0.3$ to $0.5$ for typical trap
parameters.

We thank Sandy Fetter, Tetsuro Nikuni, Reinhold Walser, and Milena
Imamovi{\mbox{\'{c}}}-Tomasovi{\mbox{\'{c}}} for
useful discussions. This work was supported by NSERC.

\appendix
\section{Collision integrals}
The two collision terms on the right-hand side of \eq{3}
are given by~\cite{Zaremba99} \bea C_{22}[f] &=& {2g^2\over (2\pi)^5
\hbar^7} \int \!\!d{\bf p}_2\int \!\!d{\bf p}_3 \int \!\!d{\bf p}_4
\nonumber \\ &\times& \delta ({\bf p}+{\bf p}_2 -{\bf p}_3 -{\bf
p}_4)\delta(\tilde \varepsilon_{p}+\tilde \varepsilon_{p_2} -\tilde
\varepsilon_{p_3}-\tilde \varepsilon_{p_4}) \nonumber \\
&\times&\left[(1+f)(1+f_2)f_3f_4-ff_2(1+f_3)(1+f_4)\right], \nonumber
\\
\label{eqa1}
\eea which is the usual Uehling-Uhlenbeck form of the Boltzmann
collision integral describing collisions between thermal atoms, and
\bea C_{12}[f,\Phi]&=&{2 g^2 n_c \over (2\pi)^2\hbar^4} \!\int \!\!d{\bf
p}_1 \!\int \!\!d{\bf p}_2 \!\int \!\!d{\bf p}_3 \nonumber \\ &\times&
\delta(m{\bf v}_c+{\bf p}_1-{\bf p}_2-{\bf p}_3)
\delta(\varepsilon_c+\tilde \varepsilon_{p_1} -\tilde
\varepsilon_{p_2}-\tilde \varepsilon_{p_3}) \nonumber \\ &\times&
[\delta({\bf p}-{\bf p}_1)-\delta({\bf p}-{\bf p}_2) -\delta({\bf
p}-{\bf p}_3)] \cr &\times&[(1+f_1)f_2f_3-f_1(1+f_2)(1+f_3)],
\label{eqa2}
\eea describing the collisional exchange of particles between the
condensate and non-condensate.

\section{Proof of Eq. 46}

We give a formal proof that there is no first-order correction to the
condensate frequencies from $\hat{\Gamma}$ in \eq{15}. From \eq{corr1}
and \eq{corr2} we can write \beq {\rm{Re}}\varepsilon_{1j} = {1 \over
2} {\int d \br w_{0j}^\dagger (\hat{M}' + \hat{M}'^\dagger) w_{0j}
\over {\int d \br w_{0j}^\dagger \sigma_3 w_{0j}}}.
\label{eqb1}
\eeq
From the explicit form of $\hat{M}'$ given in \eq{19}, we obtain
$(\hat{M}' + \hat{M}'^\dagger) = 2 i \sigma_2 \hat{\Gamma}$, where
\begin{equation}
\begin{array}{ccc}
\sigma_2 &=&
\left( \begin{array}{cc}  
	0  & -1 \\ 
	1 & 0
\end{array} \right) .
\end{array}
\label{eqb2}
\end{equation}
Then \eq{b1} becomes
\beq
{\rm{Re}}\varepsilon_{1j}  = {i \int d \br 
w_{0j}^\dagger  \sigma_2 \hat{\Gamma} w_{0j}
\over {\int d \br w_{0j}^\dagger \sigma_3 w_{0j}}}.
\label{eqb3}
\eeq
We next define $w' = \hat{\Gamma} w_{0j}$, which we expand in
the basis of the undamped eigenmodes
\beq
w' = \sum_i c_i w_{0i}.
\label{eqb4}
\eeq
Substituting this into \eq{b3} gives
\beq
{\rm{Re}}\varepsilon_{1j}  = i \sum_i c_i
{\int d \br 
w_{0j}^\dagger  \sigma_2 w_{0i}
\over {\int d \br w_{0j}^\dagger \sigma_3 w_{0j}}}.
\label{eqb5}
\eeq
We make use of the general result~\cite{Fetter72}
\bea
\int d \br w_{0j}^\dagger(\br)  \sigma_2 w_{0i}(\br)
&=& \int d\br [v_j^*(\br)u_i(\br) - u_j^*(\br)v_i(\br)] 
\nonumber \\
&=& 0 ,
\label{eqb6}
\eea
which holds for all $i$ and $j$. Using this in \eq{b5}
gives the final result \eq{realzero}.


\end{document}